\documentclass[twocolumn]{aastex62}

\usepackage{amsmath}

\received{}
\revised{}
\accepted{}
\submitjournal{ApJ}

\shorttitle{Mechanical alignment in a disk}
\shortauthors{Kataoka et al.}
\begin{document}

%\title{Millimeter-wave polarization by mechanical grain alignment in protoplanetary disks}
\title{Millimeter-wave polarization due to grain alignment by the gas flow in protoplanetary disks}

\correspondingauthor{Akimasa Kataoka}
\email{akimasa.kataoka@nao.ac.jp}

\author[0000-0003-4562-4119]{Akimasa Kataoka}
\affil{National Astronomical Observatory of Japan, Osawa 2-21-1, Mitaka, Tokyo 181-8588, Japan}

\author{Satoshi Okuzumi}
\affiliation{Department of Earth and Planetary Sciences, Tokyo Institute of Technology, 2-12-1 Ookayama, Meguro-ku, Tokyo 152-8551, Japan}

\author{Ryo Tazaki}
\affiliation{Astronomical Institute, Graduate School of Science Tohoku University, 6-3 Aramaki, Aoba-ku, Sendai 980-8578, Japan}

%
%\author{Butler Burton}
%\affiliation{National Radio Astronomy Observatory}
%\affiliation{AAS Journals Associate Editor-in-Chief}
%\nocollaboration
%\altaffiliation{Creator of AASTeX v6.2}
%\collaboration{(LaTeX collaboration)}
%\affiliation{AAS Director of Publishing}
%\affiliation{American Astronomical Society \\
%2000 Florida Ave., NW, Suite 300 \\
%Washington, DC 20009-1231, USA}
%
%\author{Jeff Lewandowski}
%\affiliation{IOP Senior Publisher for the AAS Journals}
%\affiliation{IOP Publishing, Washington, DC 20005}



\begin{abstract}
Dust grains emit intrinsic polarized emission if they are elongated and aligned in the same direction. 
The direction of the grain alignment is determined by external forces, such as magnetic fields, radiation, and gas flow against the dust grains.
In this letter, we apply the concept of the grain alignment by gas flow, which is called mechanical alignment, to the situation of a protoplanetary disk.
We assume that grains have a certain helicity, which results in the alignment with the minor axis parallel to the grain velocity against the ambient disk gas and discuss the morphology of polarization vectors in a protoplanetary disk.
%By estimating precession timescales, we find that mechanical alignment can be the dominant mechanism of determining the direction of the grain alignment in a protoplanetary disk.
We find that the direction of the polarization vectors depends on the Stokes number, which denotes how well grains are coupled to the gas.
If the Stokes number is less than unity, orientation of polarization is in the azimuthal direction since the dust velocity against the gas is in the radial direction.
%In contrast, if the Stokes number is larger than unity, the orientation is in the radial direction since the dust velocity against the gas in the azimuthal direction.
If the Stokes number is as large as unity, the polarization vectors show a leading spiral pattern since the radial and azimuthal components of the gas velocity against the dust grains are comparable.
This suggests that if the observed polarization vectors show a leading spiral pattern, it would indicate that Stokes number of dust grains is around unity, which is presumably radially drifting.

\end{abstract}



\section{Introduction} \label{sec:intro}

Millimeter-wave polarization of protoplanetary disks has been dramatically developing owing to the high-resolution and high-sensitivity observations with Atacama Large Millimeter/submillimeter Array \citep[ALMA; e.g.,][]{Kataoka16b, Kataoka17, Stephens17, Hull18, Lee18, Cox18, Sadavoy18, Harris18, Alves18,Bacciotti18,Dent19}.
The ALMA observations have revealed that the polarization of protoplanetary disks is not the straight extension from that in the star-forming regions, where grains are presumably aligned with the magnetic fields \citep[e.g.,][]{Girart06, Hull17, Maury18}.
Instead, the millimeter-wave polarization of disks is due to combination of self-scattering and alignment.
\citet{Kataoka15} have theoretically pointed out that thermal dust emission is scattered off of the dust grains themselves, and anisotropy of radiation fields gives a few percent polarization at millimeter wavelengths \citep[see also][]{Yang16a, Kataoka16a, Pohl16}.
This was confirmed in several observational studies \citep{Kataoka16b, Fernandez-Lopez16, Stephens17, Girart18, Cox18,Hull18, Lee18,Bacciotti18,Dent19}.
Furthermore, some disks show azimuthal polarized pattern, which cannot be explained by the self-scattering \citep{Kataoka17, Stephens17, Bacciotti18} but is interpreted with radiation alignment \citep{Tazaki17}.

In this letter, we introduce another mechanism of the grain alignment, which is the mechanical alignment, where gas flow may align dust grains.
The mechanical alignment was proposed by \citealt{Gold52} and has extensively discussed in the context in the interstellar medium \citep[e.g.,][]{DraineWeingartner96,DraineWeingartner97,LazarianHoang07b, Hoang18a}.
However, it has not been applied to dust grains in a protoplanetary disk.
In addition, this work is motivated both theoretically and observationally.
In the theoretical point of view,  \citealt{Tazaki17} did not treat mechanical alignment.
In terms of observations, \citealt{Yang19} have shown that the azimuthal polarization pattern of the Band 3 polarization image of HL Tau cannot be explained by the radiation alignment since it would produce a circular pattern.
This also motivates us to investigate the possibility of the mechanical alignment.

This letter is organized as follows.
We discuss the orientation of the polarization due to the mechanical alignment in Section \ref{sec:orientation}.
Since the velocity difference between gas and a dust grain depends on how well the grain is coupled with the gas, equivalently the grain size and gas density, the direction of the polarization is non-trivial.
In Section \ref{sec:discussion}, we qualitatively discuss if the mechanical grain alignment could explain previous observational results of millimeter-wave polarization.
We conclude in Section \ref{sec:conclusion}.

\section{Direction of Polarization by Mechanical Alignment} \label{sec:orientation}

The general alignment process of a dust grain in the universe is as follows \citep[see][]{DraineWeingartner96,DraineWeingartner97, LazarianHoang07a,LazarianHoang07b, Tazaki17, Hoang18a}.
A dust grain starts spinning by some external forces such as radiation.
Regardless of the external force, once the dust grain is spun up, it starts precession around one axis, such as the direction of the magnetic fields, gas flow, and radiation.
%After some random motion of the grain, it would be aligned with the precession axis.
The external torque may constantly change the orientation of the grain.
As a consequence, the grain would align with the stable precession axis.
Therefore, if we find the precession axis of the dust grain, it would be the alignment axis.
For example, in the case of the interstellar medium, the direction of the alignment is determined by the magnetic fields because the Larmor precession timescale is the shortest.
While the direction of the precession axis in a protoplanetary disk still needs a discussion, in this letter, we assume that dust grains are aligned by the gas flow due to the mechanical alignment, and focus on phenomenological discussion of the polarization vectors in a protoplanetary disk.

\begin{figure*}
\plotone{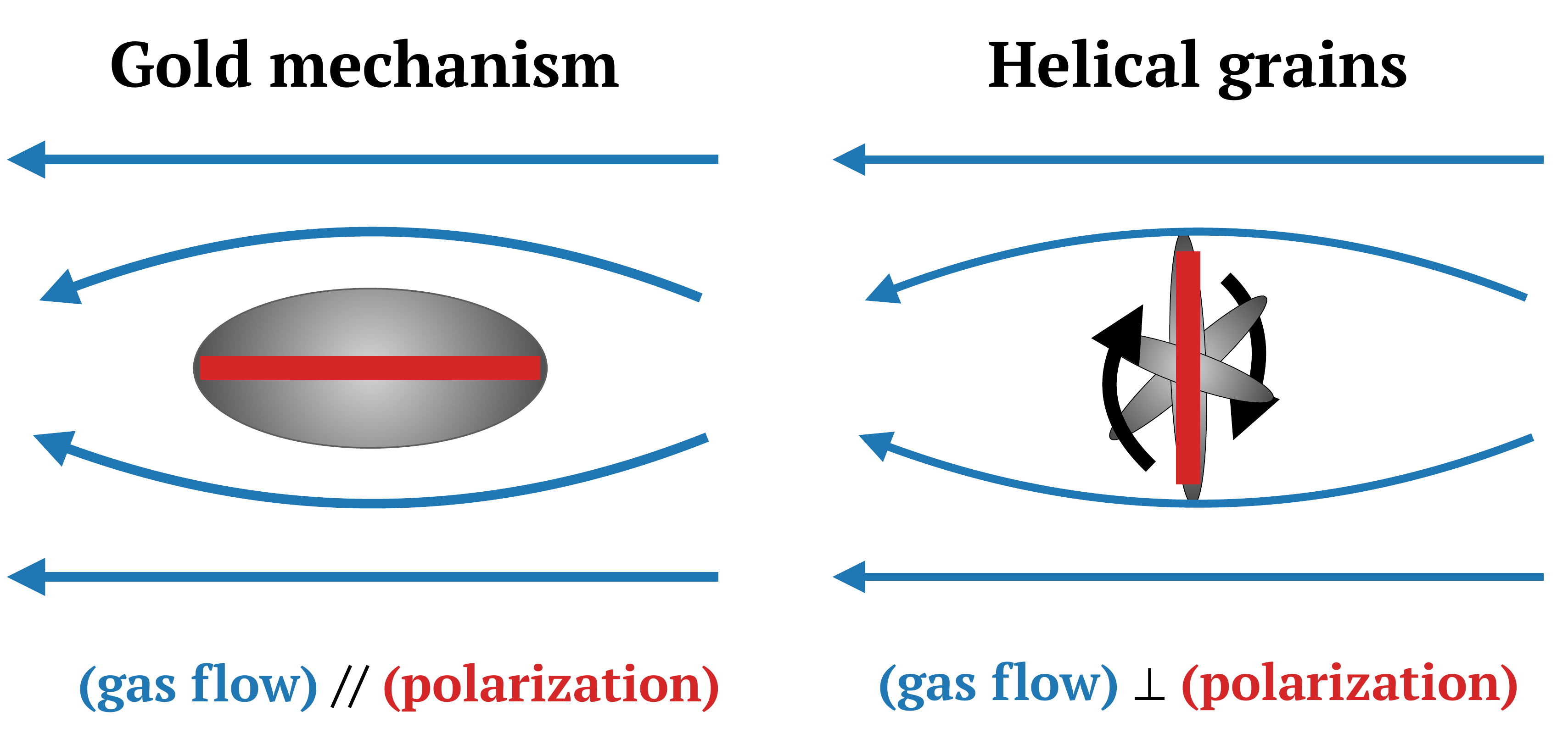}
\caption{Schematic illustration of the direction of polarization with respect to the gas velocity against a dust grain.
The left panel represents the Gold mechanism \citep{Gold52}, where the polarization orientation (red) is parallel to the velocity vector (blue).
The right panel represents the helical grain mechanism \citep{LazarianHoang07b}, where the polarization orientation (red) is perpendicular to the velocity fields (blue).
Since the gas flow in a protoplanetary disk is subsonic, the helical grain mechanism would be realized in the disk.
}
\label{fig:schematic}
\end{figure*}

Figure \ref{fig:schematic} shows the schematic illustration of the microscopic polarimetric direction of a mechanically aligned dust grain. 
The original idea has been proposed by \citealt{Gold52}, where an oblate grain is aligned with the ambient gas for its major axis to be parallel to the direction of the gas velocity against the dust grain.
In this case, the polarization vector is parallel to the gas velocity against the dust grain as shown in Fig. \ref{fig:schematic}.
However, \citealt{LazarianHoang07b} proposed that a helical grain can be more efficiently aligned with the velocity fields, where the a helical grain is aligned for the rotational axis parallel to the velocity fields.
In contrast to the Gold mechanism, as a result of the rotation, the polarization orientation is perpendicular to the velocity of the ambient gas.
The helical grain alignment works even for subsonic gas in contrast to the Gold mechanism.
Since the velocity difference between gas and dust in a protoplanetary disk is subsonic, we take the helical-grain alignment as a prior mechanism at work in a disk.

We note the relationship between helicity and polarity. 
The helicity can be either right-handed or left-handed. 
In both cases, the polarization is perpendicular to the direction of the direction of the gas velocity.
Thus, as long as the dust grains have dispersion in the helicity distribution, the dust grains have polarization due to helical-grain alignment.
Whether dust grains can obtain a sufficient helicity through the coagulation process is in question, but it is beyond the scope of this letter.

\begin{figure}
\plotone{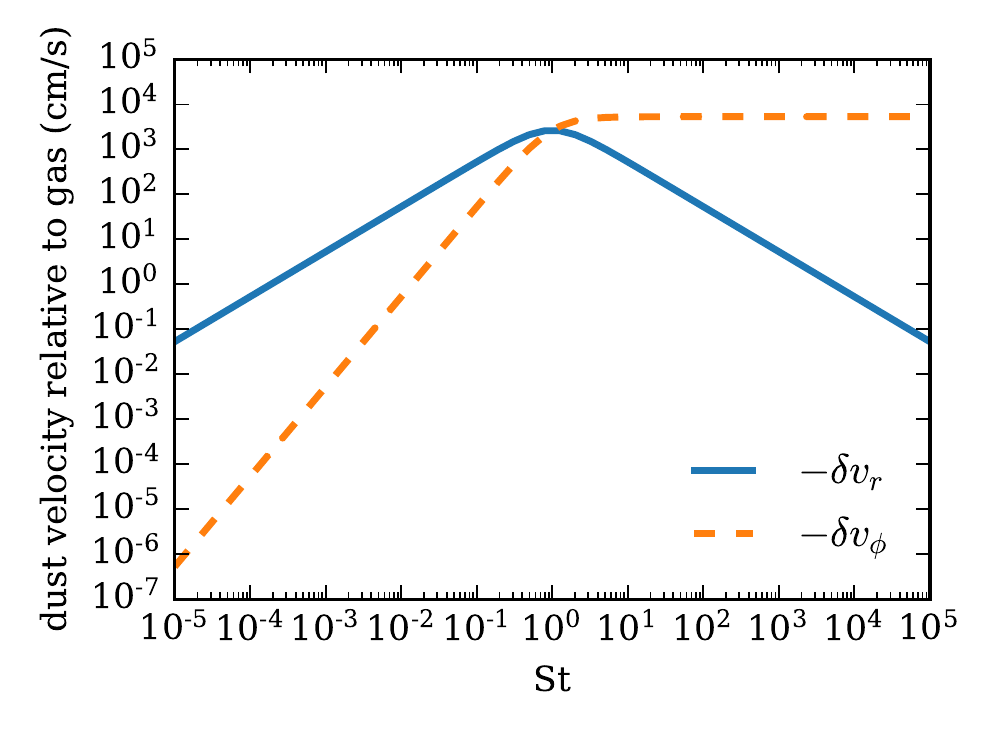}
\caption{The radial and azimuthal components of the velocity difference between gas and dust, Equations (\ref{eq:1}) and (\ref{eq:2}), in a protoplanetary disk with $\eta v_{\rm K}=53 {\rm~m~s^{-1}}$ and set $v_{\rm gas, r}=0$.
}
\label{fig:deltav}
\end{figure}

\subsection{Velocity between gas and dust} \label{sec:obs}

First, we consider the direction of the gas velocity against a dust grain in a protoplanetary disk. 
We write the radial and azimuthal component of the velocity of the dust grain as $v_{\rm dust, r}$ and $v_{\rm dust,\phi}$, and those of the gas velocity as $v_{\rm gas, r}$ and $v_{\rm gas,\phi}$.
Then, the velocity differences between gas and dust on each component, $\delta v_{\rm \phi}$ and $\delta v_{\rm r}$, are written as
\begin{eqnarray}
\delta v_{\rm \phi} &\equiv& v_{\rm gas,\phi} - v_{\rm dust, \phi} \nonumber\\
&=& - \frac{{\rm St}^2}{1+{\rm St}^{2} }\eta v_{\rm K}, \label{eq:1}\\
\delta v_{\rm r} &\equiv& v_{\rm gas,r} - v_{\rm dust, r} \nonumber\\
&=&  - \frac{{\rm St}^{-1} v_{\rm gas, r} - \eta v_{\rm K}}{{\rm St}+{\rm St}^{-1}}, \label{eq:2}
\end{eqnarray}
where St is the Stokes number, which is the dust stopping time normalized with the Keplerian timescale, and $\eta v_{\rm K}$ is the gas rotation velocity relative to the Keplerian velocity \citep[e.g.,][]{TakeuchiLin02}.
We ignore the vertical component for simplicity.
Figure \ref{fig:deltav} shows $\delta v_{\rm \phi}$ and $\delta v_{\rm r} $where we assume that $\eta v_{\rm K}=53 {\rm~m~s^{-1}}$ and set $v_{\rm gas, r}=0$ for simplicity.
The radial velocity dominates when the Stokes number is less than unity, and the azimuthal velocity dominates when the Stokes number exceeds unity.

\begin{figure*}
\plotone{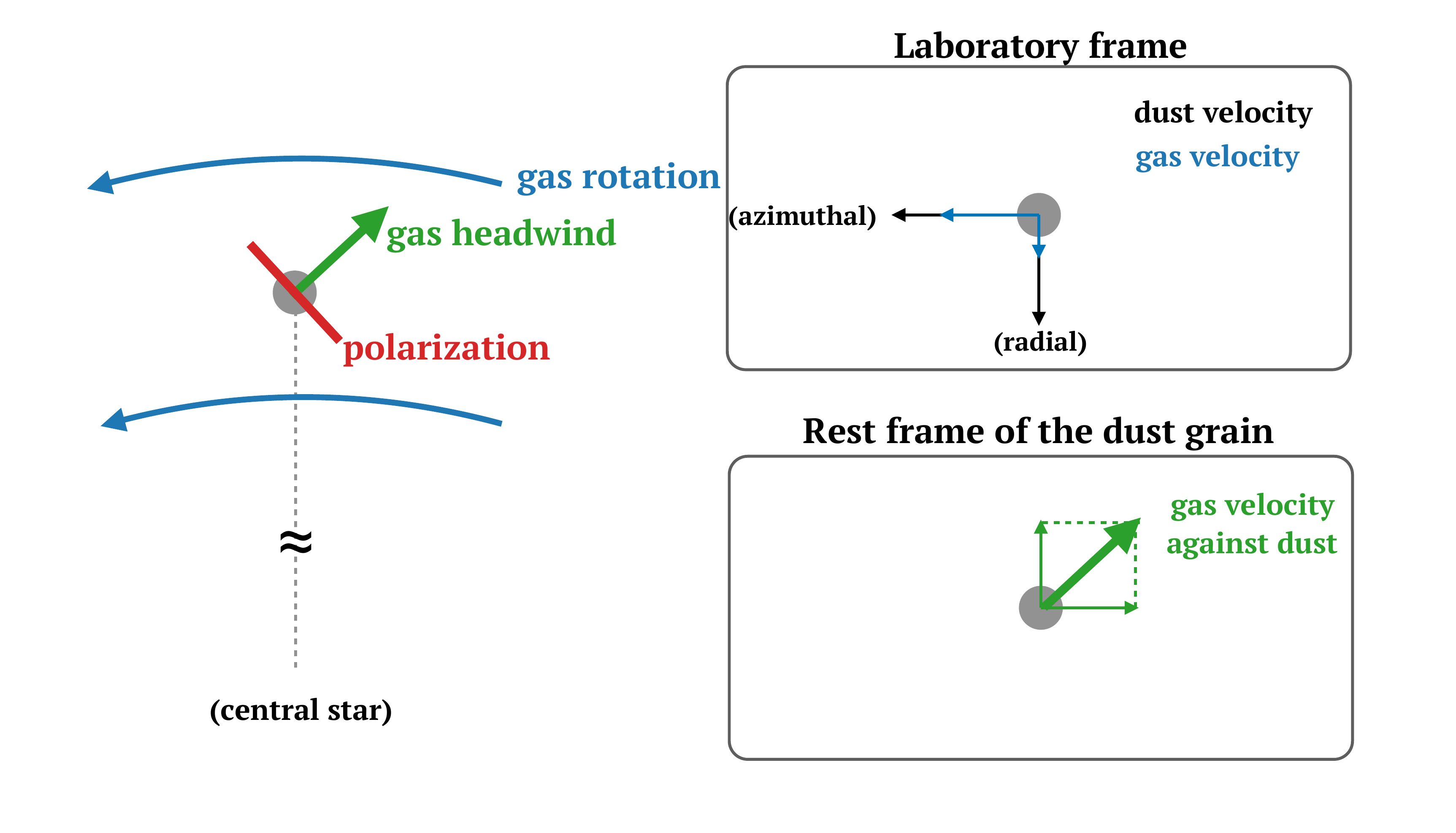}
\caption{
Schematic illustration of velocity vectors of dust and gas.
The left panel shows the location which we are considering.
We assume that the disk gas is rotating counterclockwise and the dust is feeling the headwind from the gas.
The top right panel shows the azimuthal and radial components of the gas and dust velocities.
Dust is faster than gas in rotation as well as the radial drift.
The bottom right panel shows the gas velocity on the rest frame of dust.
The polarization is perpendicular to this velocity vector, which is shown as a red line in the left panel.
}
\label{fig:deltav}
\end{figure*}

\begin{figure*}
\includegraphics[scale=0.48]{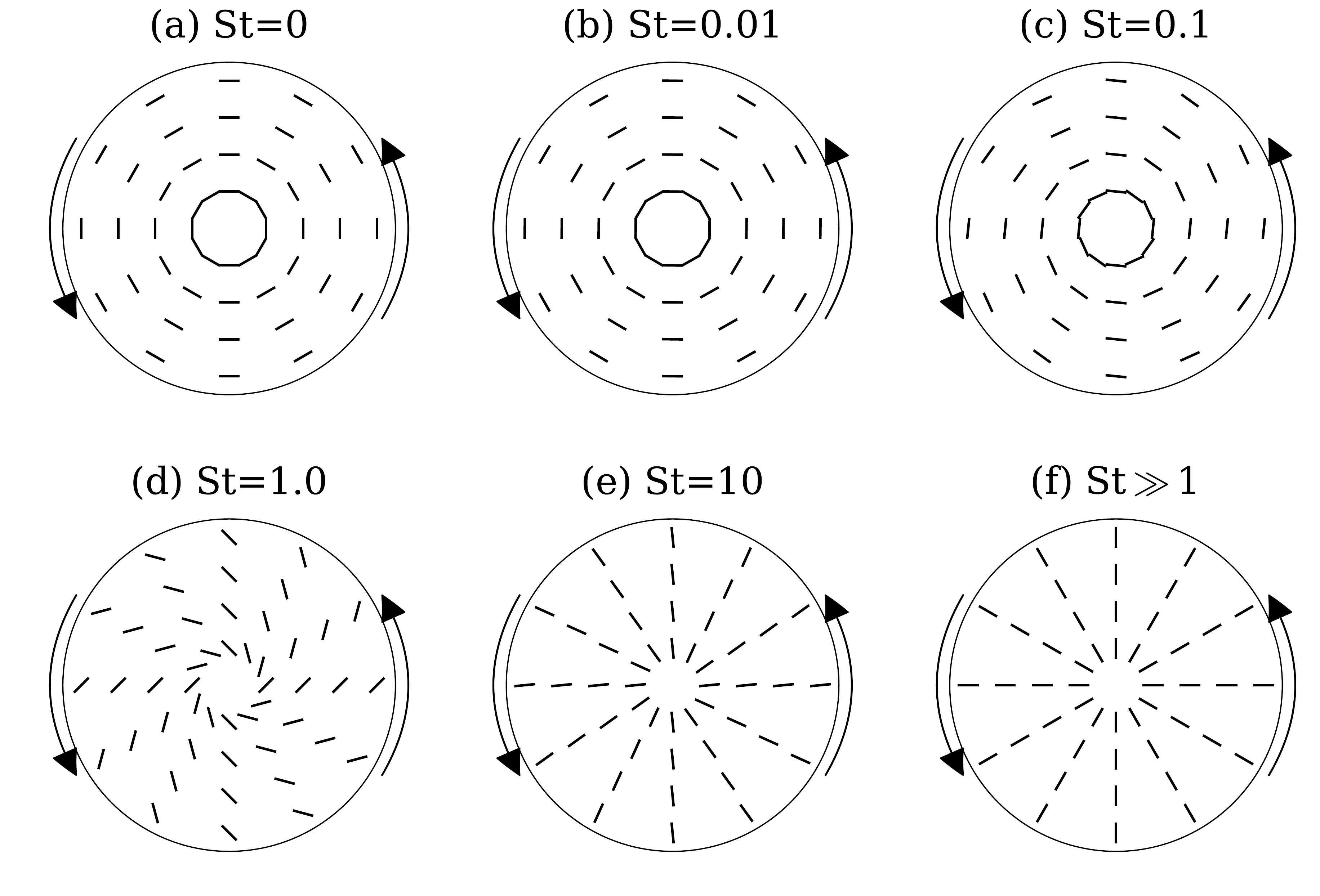}
\caption{Morphologies of the polarization vectors emitted by dust grains aligned with the gas flow due to the mechanical alignment.
Different panels show the cases of the Stokes number is $0, 0.01, 0.1, 1.0, 10$, and much greater than unity.
All disks are face-on viewed and rotating counterclockwise.}
\label{fig:vector}
\end{figure*}

Now, we are able to calculate the direction of the polarization if the dust grain is aligned by the gas flow due to its helicity. 
The polarization vector is perpendicular to the gas velocity against the dust grain. 
More generally, the polarization vectors $\bf{P}$ with a unit length can be written as
\begin{eqnarray}
{\bf P} &=& (\delta v_{\rm \phi}^2 +  \delta v_{\rm r}^2)^{-0.5} (-\delta v_{\rm \phi} {\bf e}_{\rm r}+ \delta v_{\rm r}{\bf e}_{\rm \phi} )\nonumber\\ 
&=& \frac{1}{\sqrt{1+{\rm St}^{2}}} \left({\rm St}{\bf e}_{\rm r} +{\bf e}_{\rm \phi}\right)
\end{eqnarray}
where ${\bf e}_{\rm r}$ and ${\bf e}_{\rm \phi}$ are the unit vectors in the radial and azimuthal directions, respectively.
Since we do not know the polarity, $-\bf{P}$ is also a solution for the polarized emission.

\begin{figure}
\plotone{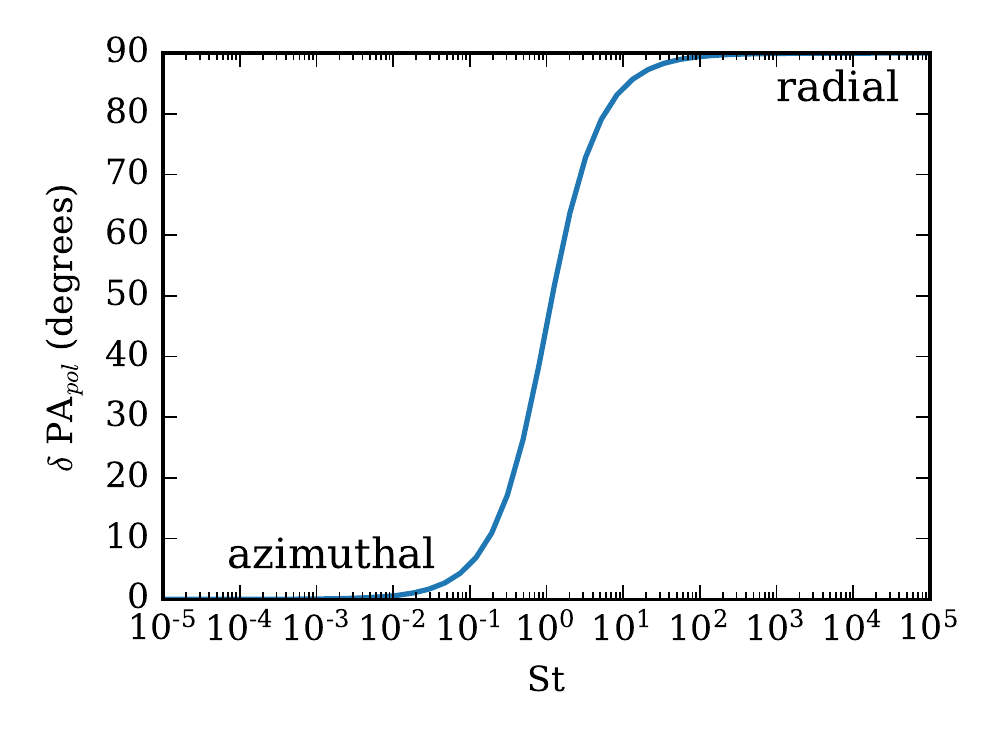} 
\caption{The difference of polarization angle from the azimuthal direction as a function of Stokes number. The expression is $\delta {\rm PA}_{pol}=\arctan({\rm St})$.}
\label{fig:deltaPA}
\end{figure}

Figure \ref{fig:vector} shows the polarization vectors of a face-on protoplanetary disk.
The six panels represent the pattern of linear polarization vectors with Stokes numbers, $0, 0.01, 0.1, 1.0, 10,$ and infinity. 
In the cases of ${\rm St}\le0.01$, the polarization is almost in the azimuthal direction.
When St becomes unity, the radial and azimuthal components of gas velocity against the dust grain become a comparable value, and thus the resultant polarization would show a spiral pattern. 
The spiral looks as a leading mode spiral, which would be a particular feature of mechanical grain alignment. 
When the Stokes number exceeds unity, the polarization vectors are in the radial direction.
The orientation of polarization is simply expressed as $\arctan({\rm St})$, as shown in Figure \ref{fig:deltaPA}.
Measurements of the orientation angle in face-on disks may help to understand the polarization mechanism.

Future measurements of the orientation angle will give insights into the mechanism that determines the maximum size of dust grains in the disks. 
The opacity of dust grains with the Stokes number larger than unity is too small to be observed. 
As shown in Figure \ref{fig:deltaPA}, the polarization orientation is in the azimuthal direction if the Stokes number is much less than unity, which is usually the case in a protoplanetary disk for dust grains whose growth is limited by radial drift or collisional fragmentation \citep[e.g.,][]{Brauer08, Birnstiel10a}.
Therefore, the mechanical alignment basically predicts the azimuthal polarization vectors, and the spiral pattern if the grains are as large as ${\rm St}\sim1$.
For instance, the maximum Stokes number of grains undergoing drift-limited growth is estimated to be ~0.1 \citep[e.g.,][]{Brauer08}.
In this case, the deviation of the polarization angle from the azimuthal direction is $\arctan (0.1) \approx 5.7^\circ$.
Therefore, if dust grains that emit polarization are radially drifting, the polarization angle is almost azimuthal but slightly different with the deviation angle of $5.7^\circ$.

\section{Discussion}\label{sec:discussion}
\subsection{Comparison with previous observations} 
Here, we discuss if the mechanical alignment is partly the origin of polarization of the observed protoplanetary disks.
The mechanisms of the millimeter-wave polarization can be first categorized into self-scattering or alignment.
One of the peculiar characteristics of scattering-induced polarization is that polarization vectors are parallel to the minor axis of the disk for an inclined disk \citep{Yang16a,Kataoka16a}, which is observed in several targets \citep{Fernandez-Lopez16, Stephens14, Girart18, Cox18,Hull18, Lee18,Bacciotti18, Dent19}.
In case of a face-on disk, if there is a ring-like structure, the morphology of the self-scattering polarization would be in the radial direction on the ring and azimuthal outside \citep{Kataoka15, Kataoka16b, Ohashi18}.

The mechanical alignment mechanism offers another interpretation for disks that show azimuthal polarization pattern, which cannot be explained by the self-scattering.
For example, VLA 1623 and DG Tau shows the self-scattering polarization at the center but the azimuthal vectors outside.
BHB07-11's polarization pattern is almost in the azimuthal direction but somewhat spiral.
HL Tau shows the self-scattering pattern at 0.87 mm but the azimuthal patter at 3.1 mm observation.
These azimuthal polarization morphologies have been interpreted as the result of the alignment with the radiative direction \citep{Tazaki17}.
However, an azimuthal pattern is also characteristic of the alignment with the gas flow as discussed in this letter.

In particular, while \citealt{Alves18} interpreted the polarization pattern of BHB07-11 as emitted by magnetically aligned dust grains, the polarization might be due to mechanical alignment.
If the grains are aligned by the gas flow due to mechanical alignment, the polarization pattern would be a leading mode in clockwise rotation if we assume the Stokes number is less than unity because the disk is rotating clockwise \citep{Alves17}.
As shown in Figure 7 of \citealt{Alves18}, the polarization vectors show the consistent pattern of the expected polarization vectors of the mechanical alignment.
The position angle deviation from the azimuthal direction is reported to be 20.2 degrees.
If we take the tangent of this, the Stokes number is derived to be 0.37.
However, this result is not deprojected to face-on disk morphology, and thus we may have overestimated the Stokes number. 
At least, since we see a certain discrepancy of polarization vectors from azimuthal direction, the Stokes number would be on the order of 0.1.

\subsection{Circular or elliptical?} 
As pointed out by \citealt{Yang19}, if the grains are aligned by a helicity-induced torque and if the minor axis of dust grains is in the radial direction, the effects of the viewing angle on the polarization pattern is non-trivial.
If grains are aligned with the direction of radiation, they would show the azimuthal pattern of polarization vectors in a face-on disk.
If the disk has a certain inclination, however, the polarization vectors are not parallel to the projected elliptical pattern but would be always normal to the center, which results in a circular pattern of polarization vectors.
If this is true, then the mechanical alignment by gas flow would also show the circular pattern for inclined disks.
Note that they argued that mechanical alignment shows elliptical pattern but this is because they assumed the Gold mechanism, which is not likely the case in protoplanetery disks because the gas flow is presumably subsonic.

\subsection{Caveat of the model} 

We discuss the morphology of polarization vectors on the assumption that dust grains are aligned by gas flow in a laminar disk.
This assumption may not be true if the disk is turbulent.
The validity of the assumption can be estimated by comparing the turnover time of the turbulent eddies of the disk gas against the dissipation timescale of the precession around the gas-drag direction \citep[e.g.,][]{LazarianHoang07b, Tazaki17}.
Especially, dust grains with small St may not be aligned by gas flow since the velocity of the disk gas against the dust grains is small.
Furthermore, even if the turbulent motion is strong enough to dealign the dust grains, anisotropic turbulent motion may produce another way of alignment \citep{ChoLazarian03}.
We leave the discussion for future work.

\section{Conclusions} \label{sec:conclusion}
In this letter, we have assumed that dust grains are aligned with the gas flow in a protoplanetary disk and discuss the resultant polarization pattern. 
Our main findings are as follows.

\begin{itemize}
\item
Since the polarization vectors are perpendicular to the headwind of the gas against a dust grain, polarization pattern depends on the Stokes number of dust grains.
If dust grains have the Stokes number smaller than unity, the polarization vectors would be in the azimuthal direction while if they have the Stokes number larger than unity, the vectors would be in the radial direction.
If the Stokes number is on the order of unity, the polarization pattern would be a leading spiral pattern.
The deviation of the polarization angle from the azimuthal direction is given by $\arctan({\rm St})$.
\item
Millimeter-polarization observations of protoplanetary disks have reveal that some disks show the azimuthal pattern of the polarization vectors, which have been interpreted by the radiative alignment.
These disks may be reinterpreted by mechanical grain alignment.
\end{itemize}

\acknowledgments
This work was initiated at the Aspen Center for Physics, which is supported by National Science Foundation grant PHY-1607611.
This work was supported by JSPS KAKENHI Grant Numbers 18K13590 and 18H05438.

\newcommand{\SortNoop}[1]{}



\end{document}